\documentclass[prb,preprint]{revtex4}

\pdfoutput=1

\usepackage{graphicx}

\begin{document}

\title{Comment on ``On the nature of magnetic stripes in cuprate superconductors,'' by H. Jacobsen \textit{et al.}, arXiv:1704.08528v2}

\medskip 

\date{May 15, 2017} \bigskip

\author{Manfred Bucher \\}
\affiliation{\text{\textnormal{Physics Department, California State University,}} \textnormal{Fresno,}
\textnormal{Fresno, California 93740-8031} \\}

\begin{abstract}
Dynamics reduces the orthorhombicity of magnetic stripes in $La_2CuO_{4+y}$. 
The measured stripe incommensuration can be used to determine the oxygen content of the sample.

\end{abstract}

\maketitle

\pagebreak

With elastic and inelastic neutron-scattering experiments on $La_2CuO_{4+y}$ Jacobsen \textit{et al.} have provided a valuable contribution to clarify the longstanding puzzle of static \textit{vs.} dynamic stripes in the pseudogap state of cuprate superconductors.\cite{1} 
The authors' main finding is a discrepancy between the incommensuration
$\widetilde{\delta_j}(\epsilon)$ of dynamic magnetic density waves (MDWs), extrapolated to vanishing energy $\epsilon$, and the value $\overline{\delta_j}$ from static MDWs,
\begin{equation}
\widetilde{\delta_j}(\epsilon \rightarrow 0) \equiv \widetilde{\delta_j^0} \ne \overline{\delta_j} \,\,\,\,\,\,\,\,\,\,\,\,\,\,\, (j = h, k) \, .
\end{equation}
Here $h$ and $k$ label the components of unidirectional MDWs,
approximately along the $Cu$-$O$ bonds in the $CuO_2$ plane,
with respect to the orthorhombic $a$ and $b$ axes.
The values are listed in Table I, reproduced from Ref. 1 (supplementary material).
The purpose of this comment is twofold: (1) To show that a \emph{modified} display of the data, Table II, provides more insight into similarities and differences of dynamic \textit{vs.} static MDWs and qualitatively suggests a plausible explanation.
(2) To determine the level of super-oxygenation $y$ from the measured (average) incommensuration $\delta_{hk}$.

\bigskip  

\begin{table}[h!]
\begin{tabular}{ |p{3.5cm}|p{2.5cm}|p{2.5cm}|p{2.5cm}|p{2.5cm}|  }
 \hline  \hline
Neutron scattering & Peak position & MDW mode & $\delta_h$ (r.l.u.) & $\delta_k$ (r.l.u.)  \\
 \hline 
Elastic & ($\overline{100}$) & static & 0.0967(5) & 0.1237(3)  \\
 \hline   
Inelastic & ($\widetilde{100}$) & dynamic & 0.1038(15) & 0.1147(8)  \\
 \hline 
 Elastic & ($\overline{010}$) & static & 0.1233(3) & 0.0950(5)  \\
 \hline 
 \hline
\end{tabular}
\caption{Incommensuration $\overline{\delta_j}$ and $\widetilde{\delta_j^0}$ ($j = h, k$) of static and dynamic magnetic density waves, respectively, in $La_2CuO_{4+y}$, expressed in orthorhombic coordinates.
Error bars are denoted in parentheses.}

\label{table:1}
\end{table}

\begin{table}[h!]
\begin{tabular}{ |p{3cm}|p{2cm}|p{2.2cm}|p{2.5cm}|p{2.5cm}|p{3.4cm}|  }
 \hline  \hline
Scattering (Peak) & & Average $\delta_{hk}$  &Deviation $\Delta \delta_h$ \ & Deviation $\Delta \delta_k$ & Orthorhombicity $\Omega$ \\
 \hline 
Elastic  ($\overline{100}$) & $\delta_h$ , $\delta_k$ = & 0.1102 & - 0.0135 & + 0.0135 & 0.12 \\
 \hline   
Inelastic  ($\widetilde{100}$) & $\delta_h$ , $\delta_k$ = & 0.1093 & - 0.0055 & + 0.0055 & 0.05 \\
 \hline 
 Elastic  ($\overline{010}$) & $\delta_h$ , $\delta_k$ = & 0.1092 & + 0.0141 & - 0.0141 & 0.13 \\
 \hline 
 \hline
\end{tabular}
\caption{Incommensuration $\overline{\delta_j}$ and $\widetilde{\delta_j^0}$  of Table I expressed in terms of their average, 
$\delta_{hk} \equiv (\delta_h + \delta_k)/2$, 
deviation from the average, 
$\delta_j = \delta_{hk} + \Delta\delta_j$ ($j = h,k$),
and orthorhombicity $\Omega \equiv |\Delta \delta /\delta_{hk}|$.}

\label{table:2}
\end{table}

\textit{1. Magnetic density waves.} The cross components of the incommensurations at the symmetry-related elastic peaks in Table I are found so close to be essentially equal, $\overline{\delta_h}(100) \simeq \overline{\delta_k}(010) \simeq 0.0959(5)$, and vice versa, $\overline{\delta_k}(100) \simeq \overline{\delta_h}(010) \simeq 0.1235(3)$. 
In contrast, a clear difference exists between the static and extrapolated dynamic values, $|\overline{\delta_h} - \widetilde{\delta_h^0}| = 0.07 $ and $|\overline{\delta_k} - \widetilde{\delta_k^0}| = 0.09 $, taken at the (100) peak, with the static values bracketing the dynamic ones. This is the surprising result of Ref. 1.

Using the average, $\delta_{hk} \equiv (\delta_h + \delta_k)/2$, and the deviation from the average, $ \Delta\delta_j$ \, ($j = h,k$),
the same data are displayed in Table II, to be viewed as $\delta_j = \delta_{hk} + \Delta\delta_j$. It now becomes obvious that the static and dynamic averages are essentially equal, $\overline{\delta_{hk}} \simeq \widetilde{\delta_{hk}^0} \simeq 0.1095$, suggesting a \emph{commonality} of static and dynamic MDWs. The average $\delta_{hk}$ would be the tetragonal approximation of the orthorhombic incommensurations. The deviations from $\delta_{hk}$ mark their orthorhombicity,
$\Omega \equiv |\Delta \delta /\delta_{hk}|$,
being considerably larger for static MDWs than for dynamic ones, $\overline{\Omega} /\widetilde{\Omega^0} \simeq 2.5$.

Why have the dynamic incommensurations $\widetilde{\delta_j^0}$ less orthorhombicity? A static MDW can be regarded a (magnetic) superlattice of lattice constants $D_j = 1/\overline{\delta_j}$. A dynamic MDW, in contrast, can be considered a standing wave of oscillating magnetic dipoles with wavelength components $\widetilde{\lambda_j^0} = 1/\widetilde{\delta_j^0}$. 
Generally, the motional aspect of dynamics promotes isotropy (here, in the $CuO_2$ plane). The oscillations of the dynamic MDWs then tend to tetragonalize their orthorhombic wavelengths $\widetilde{\lambda_j^0}$, preventing a relaxation of the magnetic dipoles to a (static) superlattice of larger orthorhombicity. This notion is corroborated by still less orthorhombicity of dynamic MDWs with higher energy $\epsilon$---albeit to a much lesser degree, $\Delta \widetilde{\Omega}/\Delta \epsilon = - 0.007$/meV (based on Fig. 3 of Ref. 1).

\bigskip

\textit{2. Oxygen content.} A frequent problem with oxygen-enriched cuprates is uncertainty about the exact level of super-oxygenation $y$. In many cases samples are characterized by the superconducting transition temperature $T_c$ instead of the value of $y$. The sample used in Ref. 1 has $T_c \simeq 40$ K, similar to the sample used by Lee \textit{et al.}\cite{1,2} A thermogravitimetric analysis of the latter sample gave an estimate of oxygen enrichment $y = 0.12 \pm  0.01$.
The \linebreak main quantity of interest is, of course, the hole doping level $p$ (per $Cu$ atom in the $CuO_2$ plane) caused by super-oxygenation $y$. No such problem occurs in the much-studied companion lanthanum cuprates $La_{2-x}Ae_xCuO_4$ ($Ae = Sr, Ba$) that are hole-doped through infravalent cation doping $x$ with a hole doping level $p = x$. In the latter materials MDWs and charge-density waves (CDWs) appear  together, called ``stripes.''\cite{3} Their incommensuration, $\delta(x) \propto \sqrt{x - x_0^N}$, depends on the cation doping $x$, diminished by the N\'{e}el point $x_0^N = 0.02$ (collapse of 3D antiferromagnetism at $T=0$).\cite{4}  

It is tempting to extend the incommensuration formula from the cation-doped to the oxygen-enriched $La_2CuO_4$ compounds.
Assuming that each enriching oxygen atom in $La_2CuO_{4+y}$ gives rise to two doped holes, $p = 2y$, the formula for MDWs, expressed in terms of oxygen enrichment and orthorhombic coordinates (but in tetragonal approximation) becomes,
\begin{equation}
\delta_{hk}(y)  = \frac{1}{4}\sqrt {2y - x_0^N}  \; .
\end{equation}
Solving for $y = 8\delta_{hk}^2 + x_0^N/2$ and using the average incommensuration $\delta_{hk} = 0.1095$ r.l.u. gives $y = 0.105 \pm 0.005$, comparable with the estimate by Lee \textit{et al.}, $y = 0.12 \pm 0.01$.

\bigskip  \bigskip

\centerline{ \textbf{ACKNOWLEDGMENTS}}

\noindent I thank Sonja Lindahl Holm and Henrik Jacobsen for literature links and information about the oxygen content of the sample.


\begin{thebibliography}{4}

\bibitem{1} H. Jacobsen, S. L. Holm, M.-E. L\u{a}c\u{a}tu\c{s}u, M. Bertelsen, M. Boehm, R. Toft-Petersen, J.-C. Grivel, S. B. Emery, L. Udby, B. O. Wells, and K. Lefmann, ``On the nature of magnetic stripes in cuprate superconductors'', arXiv:1704.08528v2

\bibitem{2} Y. S. Lee, R. J. Birgeneau, M. A. Kastner, Y. Endoh, S. Wakimoto, K. Yamada, R. W. Erwin, S.-H. Lee, and G. Shirane, Phys. Rev. B \textbf{60}, 3643 (1999). 

 \bibitem{3} J. M. Tranquada, Physica B \textbf{407}, 1771 (2012).
 
\bibitem{4} M. Bucher, ``Universality of density waves in \textit{p}-doped ${La_2CuO_4}$ and \textit{n}-doped ${Nd_2CuO_{4+y}}$'', arXiv:1702.05364

\end{thebibliography}
\end{document}